\documentclass[11pt,a4paper]{article}
\usepackage[margin=2.4cm]{geometry}
\usepackage[T1]{fontenc}
\usepackage{lmodern}
\usepackage{microtype}
\usepackage{booktabs}
\usepackage{tabularx}
\usepackage{amsmath}
\usepackage{xcolor}
\usepackage{graphicx}
\usepackage{enumitem}
\usepackage{longtable}
\usepackage[hidelinks]{hyperref}
\setlist{itemsep=2pt,topsep=3pt}

\newcommand{\usmap}{\texttt{.usmap}}

\title{Marauder: Recovering Unreal Engine type mappings from game files alone}
\author{Anastasios Antoniadis\\[1mm]\small Sourcemined (\url{https://sourcemined.com}) \quad \texttt{anantoni@sourcemined.com}\\
\small ORCID \href{https://orcid.org/0009-0004-1605-967X}{0009-0004-1605-967X}}
\date{November 2026}

\begin{document}
\maketitle

\begin{abstract}
Recent Unreal Engine games store their assets with \emph{unversioned} property serialization: a cooked object lists its property values without names or types, and a reader needs the class layouts (a ``mapping'' or \usmap{} file) to decode them.
Today mappings are obtained by injecting a dumper into the running game.
We describe a generator that builds the mapping from the game's files alone: the executable's compiled reflection tables, the cooked packages, and configuration files.
It does not run or inject into the game.
The main test is real loading. On every package of 121 supported games (510,503 packages, UE 4.19 to 5.7), the CUE4Parse asset library fully loads exactly the same packages with the generated mapping as with a runtime dump: 509,297 with each, so recall against the runtime dump is 100\%, and no package loads with only one of the two. A package counts as fully loaded only if every export deserializes without error, ends at its exact size, and no checked nested read is skipped; a negative control shows that this test detects damaged maps.
On 99.966\% of these packages the decoded data is the same up to the letter case of object keys, and every remaining difference is letter case of names, which the engine compares case-insensitively. The 1,206 packages that do not load fail with both maps, for reasons no mapping can fix: reader limits, a limit of our test harness, editor and test classes that are not in the shipped game, plugin serialization, reader support for hair assets added upstream later, and one benchmark artefact.
We then fixed the reader and reloaded every package of all 121 games with both maps. With our CUE4Parse fixes, current upstream code and a fix of our test harness, 961 more packages fully load (899 from the reader changes, 62 from the harness fix), with the same load result for both maps and the same data up to letter case, and none of the 509,297 packages that loaded before is lost. 99.952\% of all packages fully load with either map, and 99.971\% once the 99 packages whose classes no mapping can describe are left out.
At the field level, against runtime dumps of 97 supported games (UE 4.19 to 5.7), the latest generator gets 99.923\% of schema fields right on the 66 development games and 99.886\% on a 31-game hold-out; 19 of the 97 maps match the runtime dump exactly.
On a blind, frozen set of 24 new games, scored once per version with the generator pinned before the gold was opened, v10 scored 99.843\% and v11 99.867\%; 94\% of the remaining v11 errors are one separate class, Animation Blueprint structs, where a name-keyed reference holds one of several valid copies chosen by runtime state, and which the generator deliberately leaves out. Without that class, v11 is at 99.989\% to 99.992\% on all three sets.
On large modded games without a runtime reference, the v11 map lets 91.4\% of sampled packages load in one game (none load without a map), and a later version reaches 99.7\% there (100\% with a reader fix); the method fails on a custom engine build and on a protected executable. The scope is unencrypted games on standard engine builds.
\end{abstract}

% ---------------------------------------------------------------
\section{Introduction}

Modding tools for Unreal Engine games read cooked asset packages: to list items, inspect level data, view textures and meshes, or patch values.
Older engine versions wrote each property with a tag (name, type, size), so a reader could parse an object without knowing its class.
Recent versions commonly cook with \emph{unversioned} property serialization: an object stores a compact header saying which properties of its class are present, followed by their values in schema order, with no names or types~\cite{ue-unversioned}.
A reader therefore needs the exact class and struct schemas the game was built with: property names, their order, array dimensions and full type trees, plus enum names and entries.

The community format for this information is the \usmap{} mapping file, popularised by UnrealMappingsDumper~\cite{umd} and now read by CUE4Parse and FModel~\cite{cue4parse,fmodel} and UAssetAPI~\cite{uassetapi}.
Without a mapping, CUE4Parse refuses to parse unversioned packages.
Every public way we found to obtain a mapping inspects the game \emph{while it runs}: a DLL injected into the process (UnrealMappingsDumper, UE4SS, Dumper-7~\cite{umd,ue4ss,dumper7}), or a full memory image of the running process (jmap~\cite{jmap}).
Section~\ref{sec:related} reviews these tools.
This dependence on the running game has costs that show up in the dumpers' public issue trackers.
A mapping changes when class layouts change, and the reader's author advises regenerating it after each game update~\cite{fmodel-regen}.
Dumpers themselves break on updates: most such reports are fixed within days~\cite{d7-355,ue4ss-1483}, but some stay broken for weeks or months~\cite{ue4ss-803,d7-407,d7-358}, and new engine versions can be unsupported by release builds for a while~\cite{ue4ss-1198,umd-77}.
(Of the 166 closed issues opened on one dumper's tracker since July 2025, the median time to close was 3 days and 58 took more than 30 days; these include requests that are not update breakage.)
Games with anti-cheat block the injection, and maintainers advise against injecting into them~\cite{d7-593,ue4ss-1481,ue4ss-782}.
Community-shared mappings fill part of the gap, but the largest public archive covers about 80 games, one file per game version, and requests for missing games go unanswered~\cite{umd-archive,archive-req}.
A user who only wants to read files they already own must still launch the game, and keep doing so after every update.

This paper describes a generator that produces the mapping from the user's own game files only: the shipping executable read as data, the cooked packages (pak or IoStore containers, or loose files), and the shipped configuration files.
It never executes or injects into the game. Encrypted archives are refused by default; users may supply their own key; the tool does not obtain or break keys.

What a user needs is that their asset reader loads their packages, so that is our main evaluation.
With CUE4Parse, the generated map fully loads every package that a runtime dump fully loads, on every package of 121 supported games (509,297 of 509,297; 510,503 packages in total), and the decoded data differs only in the letter case of names (Section~\ref{sec:fullload}).
The packages that neither map loads are reader limits (and, for 62 of them, a limit of our test harness), not map errors: with our fixes for five root causes in the reader and current upstream code, 899 more packages load with both maps, and a fix of our test harness makes 62 more load.
Field-level accuracy against the runtime dumps is the secondary measure: it shows where the map still differs from the dump, mostly in Animation Blueprint structs, which the reader takes from the package, and in letter case, which the reader ignores (Section~\ref{sec:realuse}).
The contributions are:
\begin{itemize}
\item a full-load evaluation of every package of 121 supported games with CUE4Parse under the generated map and under a runtime dump, with a strict definition of a loaded package, separate runs per map, a negative control, a classification of every shared failure, and a rerun with reader fixes for the main failure causes (Section~\ref{sec:fullload});
\item a method that reconstructs the schema from the compiled reflection tables in the executable, adds the per-asset structs that the Animation Blueprint compiler generates, and recovers the display spelling of names from a model of the engine's name registration order (Section~\ref{sec:method});
\item a benchmark of runtime dumps with a frozen development/hold-out split and a blind, freeze-before-scoring fresh set, with a field-level metric and an honest record of every look at held-out data (Section~\ref{sec:bench});
\item results per version, per rule (including rules that were dropped), per engine version, and on the blind set (Sections~\ref{sec:results}--\ref{sec:fresh});
\item a sampled real-use evaluation with a targeted negative control, and a load test on large modded games that have no runtime reference (Section~\ref{sec:realuse});
\item a measurement of how much the runtime reference varies across cold launches, and which of our errors that variation explains (Section~\ref{sec:moves}).
\end{itemize}

% ---------------------------------------------------------------
\section{Background}
\label{sec:background}

\paragraph{Packages and exports.} A shipped (``cooked'') Unreal game stores its content in package files, usually bundled into large container files (pak or IoStore). Each package holds one or more serialized objects, called exports: a texture, a mesh, a level, or a Blueprint (a class defined in the editor rather than in C++). Loading a package means deserializing every export; the mapping tells the reader which fields each export stores.

\paragraph{Reflection and registration.}
Unreal's header tool generates, for every reflected class, struct and enum, a set of constant parameter records: one record per type (name, flags, pointers to its property list and to the function that constructs its super type) and one per property (name, flags, offset, type-specific extras such as an inner property, an enum or a struct)~\cite{ue-source}.
At startup the engine walks registration tables and builds the runtime type objects from these records.
A runtime dumper later walks those objects and writes the \usmap.

\paragraph{The mapping format.}
A \usmap{} holds a name table, a list of enums (name and entries) and a list of structs (name, super name, property count, and per property its schema index, array dimension, name and recursive type).
Two maps can be compared record by record.

\paragraph{Names and case.}
Engine names (FNames) are interned in a global table.
In shipping builds comparison is case-insensitive, but each entry keeps the spelling with which it was \emph{first} registered.
If engine code registers \texttt{position} before the reflection data registers a property \texttt{Position}, the property is displayed, and dumped, as \texttt{position}.
Case therefore depends on everything that registers names before reflection runs: static initializers, configuration parsing, plugin start-up code.
This is the main source of ``casing'' errors below.
Readers that match names case-insensitively are not affected, but tools that look properties up by exact spelling are.

% ---------------------------------------------------------------
\section{Method}
\label{sec:method}

The generator is a pipeline of stages.
Each stage only adds or changes a record when its evidence is unambiguous; otherwise it \emph{abstains} and leaves the previous output.
The inputs are the executable (as bytes), the cooked packages and the configuration files.

\subsection{Schema from compiled reflection tables}
For engine versions that compile reflection into constant parameter records (UE 4.17 and newer, according to the engine source; our benchmark has UE 4.19 and newer), the generator locates the class, struct, enum and property parameter records in the executable's data sections, follows their pointers and decodes each property's type tree using the record layouts of the respective engine version, taken from the public engine source~\cite{ue-source}.
Where a build leaves part of a record to a start-up initializer, it recovers what the data still holds: for example the entry array of a partly initialized enum record, bounded by the next record, the first non-identifier pair, or a change of qualifier.
Where data alone does not link a property to the enum it uses, it searches a short window (at most 64 bytes) at the start of the enum's construction function for an address operand equal to a recovered record; the instruction semantics are not used.
Builds older than UE 4.17 in our set (4.13--4.16) construct every type in code and have no such tables; recovering them would require decompiling game code, so they are out of scope by design.

Several engine rules that add or rename entries at registration time are modelled directly from the engine source:
numbered names (an integer suffix stored separately from the string),
automatic enum maxima (\texttt{<prefix>\_MAX}, with the prefix computed case-sensitively from the stored entries and the bit-flag form that changed in 5.5),
and intrinsic classes that only builds with assertions register, detected from assertion strings in the executable.

\subsection{Per-asset structs from cooked packages}
Animation Blueprints compile two structs per asset (constant and mutable data).
They are not in the executable; they are exported by the asset's cooked package and become part of the runtime type set when the asset loads.
The generator reads these exports from IoStore (zen) packages for UE 5.0--5.7, from pak files, and from loose cooked files.
It parses the struct header and serialized field list, and resolves references through the package import map and the global script-object table.
A parse must end exactly at the export's cooked size, or the record is dropped.
A child Animation Blueprint has no export of its own; at load it gets an empty struct whose super is the parent's, and the generator models this.

Which copy counts is harder.
Every Animation Blueprint has its own copy of each of the two structs, under the same two names and with different fields.
The runtime dump contains every loaded copy, in object-array order; a mapping reader, and therefore our scorer, keys records by name, so the copy that comes last in the dump is the one that counts.
Which copy is last depends on how many freed object slots the engine had when each asset loaded (Section~\ref{sec:animclass}), which is runtime state that the cooked files do not hold.
The generator computes the start-up set (the hard package closure of the configured game instance, the default map and, if needed, the default game mode) and emits a record only when every copy of that struct in the set is identical.
Otherwise it abstains (Section~\ref{sec:abstain}).

\subsection{Name spellings from a registration-order model}
To decide the displayed spelling of each name, the generator orders all name registrations it can see, and the first one wins.
The sources, in the order the final model uses, are:
(1) type names from the reflection records;
(2) names hard-coded in the engine's name table;
(3) names passed as literals to the name constructor from the C runtime's static-initializer table, in table order;
(4) configuration keys, in the load order of the engine's configuration hierarchy (base, engine, project and platform layers);
(5) native function names, registered when compiled-in classes are processed.
Struct, class and enum records follow the registration tables rather than address order.
For source (3) the generator reads, from each static-initializer callback and its direct callees, only literal string operands and direct call targets; it does not decode function logic, follow data flow, or reconstruct control flow.
This one step is what moved casing the most (Section~\ref{sec:results}): earlier versions knew 400--2,000 static names per game; the final model knows 2,000--16,000.

\paragraph{Spelling evidence from content and from similar builds.}
Two weaker sources fill gaps.
Spellings found in cooked name tables and the asset registry are used only for names with no executable evidence, because measured trials showed that letting content override the executable breaks more than it fixes.
For engine-internal names that code creates before reflection, a nearest-neighbour rule compares the target with reference builds (by Jaccard similarity of record names, so engine version and plugin set dominate) and respells a name only if at least two of the three most similar references agree on one other spelling, none kept the generated spelling, no near-tied reference kept it, and the new spelling occurs as a string in the target executable.

\subsection{Abstention rules}
\label{sec:abstain}
Abstention is the main design choice.
A missing record costs a few fields; a wrong copy of a large per-asset struct, or a wrong spelling applied to a common name, can cost hundreds.
The rules are:
\begin{itemize}
\item \textbf{Animation Blueprint copies:} emit only when all start-up copies are identical; never guess between differing copies.
\item \textbf{Reference spellings:} unanimous neighbours, at least two votes, no dissent within the similarity band, spelling present in the executable.
\item \textbf{Content spellings:} only for names absent from executable strings.
\item \textbf{Static-name call targets:} in builds whose name constructor takes no find-type argument (5.7), use the two dominant call targets; with no clear lead, abstain.
\item \textbf{Partial records:} recover entry counts only when bounded by data; otherwise leave the record as is.
\item \textbf{Inputs:} encrypted archives are refused unless the user supplies a key (the tool does not obtain or break keys); the encrypted benchmark games were excluded.
\end{itemize}

% ---------------------------------------------------------------
\section{Benchmark and protocol}
\label{sec:bench}

\paragraph{Reference (gold).}
For each game we captured a runtime mapping with the UE4SS mapping dumper~\cite{ue4ss} with Blueprint-generated types excluded (\texttt{DumpUSMAP(false)}~\cite{ue4ss-dumpers}), offline with network access denied, using bounded launches, and accepted a dump only when the last two captures produced the same schema.
We call this the \emph{runtime gold}; Section~\ref{sec:moves} discusses why it is a sample rather than ground truth.

\paragraph{Splits.}
The benchmark has 108 games: 75 development and 33 hold-out.
Six games with encrypted archives are excluded from scoring but kept in the strict-exact denominators of early versions, and five UE 4.13--4.16 games are out of scope (Section~\ref{sec:method}).
The supported, scored set is therefore 66 development and 31 hold-out games (97 in all; strict-exact ratios below are over the 103 supported games including the six encrypted ones, as in the source reports).
All rules were designed on development gold only.
For the reference-spelling rule, development games were scored leave-one-out, and hold-out games used only development references.

\paragraph{Looks at the hold-out.}
The hold-out was meant to be run once per version.
It was not always: it was run once in v9 with rules frozen, then again after a guard was added in response to that run; once in v10; and three times in v11, where two candidate orders failed and a third, designed after the second look, was kept.
By the end of v11 the hold-out had been seen six times, so it is only weak confirmation.
Scoring the released configuration of the spelling rule added three more hold-out scorings after the research numbers were final (Section~\ref{sec:avail}): a first run whose per-game totals exposed two faults, an intermediate run, and the final run. The table and version lists use development data only, but the fixes were prompted by the first run, so the hold-out figure for the released configuration is weak confirmation, like the other kept rules designed after a look.
The same three runs also scored the fresh set. The spelling-table and version-marker files were rewritten after the intermediate run and before the final one; our records do not show whether the intermediate hold-out total was read before that change. The final run scored both the leave-one-out and the shipped table, which give the same hold-out and fresh totals. A last rescoring of the maps written by the released command line gave the same totals.
We report every look.

\paragraph{Fresh blind hold-out.}
To get a clean test, a fresh list was frozen before any download.
Candidates were 165 games from a pool of titles never downloaded, captured or seen by any generator run and not in either split.
They were ranked by a salted SHA-256 hash of their store identifier; the first 24 that met the criteria form the set, with a 36-game reserve, and games that turned out encrypted, non-Unreal, older than UE 4.17 or without a stable dump were to be skipped in rank order.
The protocol then was:
(i) download in rank order;
(ii) capture gold;
(iii) commit the gold and its SHA-256 manifest (``freeze'');
(iv) pin the generator to a fixed commit (the v10 release) in a clean worktree;
(v) before scoring, check on four already-known games that this end-to-end chain reproduces the benchmark's cached maps exactly (it did: 0 struct and 0 enum differences);
(vi) build all 24 maps, and only then open the gold, verifying its hashes against the manifest.
No rule was changed after scoring.
The v11 generator, frozen before the v10 fresh score was known, was later run once on the same frozen gold in the same way.

\paragraph{Documented deviations.}
Two departures from the frozen rule happened, both before the freeze and both recorded at the time.
First, only about one candidate in four was an Unreal game, so the listed reserve ran out; selection continued in the same frozen hash order, and the full ranking was recomputed and checked to match the frozen list on its first 60 entries.
Second, six games could not be downloaded (the store client reported a failed state after the update job); ``download failed'' is not one of the four frozen skip reasons, but the games could not be obtained and were skipped in rank order.
In total 101 rank positions were examined: 24 accepted (positions 4 to 101), 70 non-Unreal, 6 download failures and 1 game without a stable dump after three attempts.
One accepted game first hit a tooling timeout under server load and was retried; that is not a skip reason.

\paragraph{Metric.}
We score \emph{records} and \emph{fields} against gold.
Records are types and enums; the denominator is the union of generated and gold names, so both missing and extra records count, and a record is right only if it matches exactly.
For each type there are 2 header fields (super, property count) and 4 fields per property position (name, full type tree, index, array dimension), with properties aligned by position; each enum has 1 field per entry, compared as a multiset.
The \emph{case-insensitive} variant folds ASCII case on all names first.
\emph{Strict exact} means the whole map equals gold.
With the v11 maps the development set has 7,398,988 scored fields, the old hold-out 3,309,942 and the fresh set 2,778,196 (the record denominator is a union, so totals move slightly with the generated map).
A ``game worse'' is a game whose count in any difference class increases; the release rule was that no rule may make any development or hold-out game worse.

% ---------------------------------------------------------------
\section{Results}
\label{sec:results}

\subsection{Headline}
Table~\ref{tab:headline} gives the latest version (v11) on both splits.
Field accuracy is 99.923\% on development and 99.886\% on the old hold-out; ignoring case it is 99.933\% and 99.897\%.
Strict exact is 15 of 66 and 4 of 31.

\begin{table}[t]
\centering\small
\caption{v11 accuracy on the supported games (UE $\ge$ 4.17, not encrypted). Fields and records as defined in Section~\ref{sec:bench}.}
\label{tab:headline}
\begin{tabular}{lrrrrrr}
\toprule
Split & Games & Strict & Fields & Fields (no case) & Records & Wrong fields\\
\midrule
Development & 66 & 15 & 99.923\% & 99.933\% & 99.854\% & 5,696\\
Old hold-out & 31 & 4 & 99.886\% & 99.897\% & 99.836\% & 3,774\\
\bottomrule
\end{tabular}
\end{table}

All v11 numbers in this paper come from full pipeline runs: every stage was rerun from the game files on all 121 scored games, and the maps were scored with the unchanged scorer.
The first v11 development report had applied the change in display order to cached v10 maps; the full run reproduced its development and hold-out totals exactly (5,696 and 3,774 wrong fields).

\subsection{Version history}
Table~\ref{tab:history} shows how each version moved the numbers.
v7 and v8 fixed specific registration rules; v9 introduced the field metric and the reference-spelling rule; v10 added the Animation Blueprint reader; v11 the registration-order model.
v12 was the copy-choice experiment of Section~\ref{sec:animclass} and was not kept; v13 contains the naming changes of Section~\ref{sec:modded}.
Casing-only type fields on development fell from 5,076 (v8) to 672 (v11).

\begin{table}[t]
\centering\small
\caption{Progress by version. Strict-exact over the 103 supported games (6 encrypted included in the denominator); v6--v7 over all 108. Fields are field accuracy on development / old hold-out.}
\label{tab:history}
\resizebox{\linewidth}{!}{\begin{tabular}{lcccc}
\toprule
Version (main change) & Strict & Fields dev / hold & Wrong dev / hold & Casing fields dev / hold\\
\midrule
v6 (baseline) & 9/108 & -- & -- & 5,250 / 1,988 \\
v7 (numbered names, maxima, UE 4.19 class) & 12/108 & -- & -- & 5,174 / 1,936\\
v8 (partial enum records, maxima) & 13/103 & 99.836 / 99.839 & 12,118 / 5,343 & 5,076 / 1,910\\
v9 (reference spellings, intrinsics) & 14/103 & 99.865 / 99.878 & 9,954 / 4,030 & 2,968 / 619\\
v10 (Animation Blueprint structs) & 15/103 & 99.892 / 99.878 & 8,016 / 4,022 & 2,968 / 619\\
v11 (registration order) & 19/103 & 99.923 / 99.886 & 5,696 / 3,774 & 672 / 375\\
\bottomrule
\end{tabular}}
\end{table}

The v6 and v7 casing counts were measured before the v9 metric existed and include the five out-of-scope games; they are not directly comparable with later rows.

\subsection{Per-rule ablation, including dropped rules}
Table~\ref{tab:ablation} lists every rule that was measured, cumulatively where it was kept, with the reason for dropping the others.
The rule ``no game worse on either split'' removed several rules that improved the aggregate.

\begin{table}[p]
\centering\small
\caption{Per-rule results (wrong fields; lower is better). ``Worse'' counts games that got worse in any difference class. Kept rules are cumulative within their version.}
\label{tab:ablation}
\begin{tabularx}{\linewidth}{@{}lXrrrrl@{}}
\toprule
Ver. & Rule & Dev wrong & Dev worse & Hold wrong & Hold worse & Status\\
\midrule
v8 & baseline & 12,118 & -- & 5,343 & -- & \\
v9 & Startup configuration key order (alone on v8) & 12,073 & 0 & 5,352 & 2 & dropped\\
v9 & Reference spellings, frozen version & 9,651 & 0 & 4,064 & 1 & replaced\\
v9 & Reference spellings with similarity-band guard$^{a}$ & 9,966 & 0 & 4,030 & 0 & kept\\
v9 & + intrinsic classes & 9,954 & 0 & 4,030 & 0 & kept\\
v10 & + identical Animation Blueprint copies & 8,016 & 0 & 4,022 & 0 & kept\\
v10 & + agreement rule (last created = only closest copy) & 6,480 & 0 & 4,442 & 5 & dropped\\
v10 & always last created (measured only) & 5,965 & 15 & 3,889 & 13 & not used\\
v11 & new static-name list, v10 order & 6,472 & 0 & -- & -- & kept\\
v11 & + static names before type and function names & 5,743 & 0 & look 2$^{b}$ & 1 & replaced\\
v11 & + configuration keys & 5,698 & 0 & look 2$^{b}$ & 1 & kept\\
v11 & + native registration order & 5,696 & 0 & look 2$^{b}$ & 1 & kept\\
v11 & + enums before structs & 5,698 & 2 & -- & -- & dropped\\
v11 & final order: type names first$^{a}$ & 5,696 & 0 & 3,774 & 0 & kept\\
\bottomrule
\end{tabularx}
\raggedright\footnotesize
$^{a}$Designed after a hold-out look; its hold-out number is weak confirmation.
$^{b}$Hold-out look 2 ran these three cumulative candidates in one run (3,798, 3,795 and 3,795 wrong fields), each making one game worse; the final order replaced the ``static before type names'' placement. Look 1 (static names as a post-pass) gave 3,770 and strict 5/31 but also made one game worse and was dropped.
\end{table}

Other trials that were measured and rejected on development, before any hold-out run:
letting asset-registry spellings override executable spellings dropped strict exact from 6 to 1 of 75 (v7);
an unscoped first version of the enum-maximum rule broke 796 enum entries across 66 games (v8);
UTF-16 literal spellings fixed 70 fields and broke 5,539 (v9);
content spellings applied without the executable condition fixed 3, 5 and 14 and broke 76, 77 and 136 (redirects, asset registry, registry name table);
package name tables on seven UE4 games fixed 8 and broke 71;
and applying the v11 order delta only to names no later stage had touched scored 6,559 against 6,472.

The pattern is consistent: evidence from content and from loose string order is plentiful but unreliable, and every rule that used it without an executable-side condition did more harm than good.

\subsection{Error classes}
After v10, 4,914 of the 8,016 wrong development fields (61\%) and 3,396 of 4,022 hold-out fields (84\%) were missing Animation Blueprint structs where the rule abstained (27 development and 15 hold-out games).
After v11 the class is unchanged: 4,914 of the 5,696 wrong development fields and 3,396 of the 3,774 hold-out fields, because v11 changed only spellings (Table~\ref{tab:causes}).
Section~\ref{sec:animclass} explains why we report it separately.
Of the 672 development casing type fields (plus 30 casing enum entries), the largest groups are a tooltip-key spelling in ten games (80 fields, probably a struct built early by a module start-up call), game-module names created by runtime code, type-name strings built from static strings rather than passed as literals, and names registered by module start-up code such as input-key initialization.
None of these is visible to a literal-operand model.

\subsection{Animation Blueprint structs: a separate error class}
\label{sec:animclass}
The two struct names involved, the generated constant-data and mutable-data structs, are not built at load time.
They are cooked exports of each Animation Blueprint package, created by the Blueprint compiler in the editor and only read in at load; a child Animation Blueprint links its struct to the parent's, and one without its own export falls back to the parent's.
The generator reads them correctly: on development gold (34 UE5 games that contain such records), 311 of 313 gold copies equal a copy read from the game's packages, and the other 2 come from a package the scan did not read.
The v10 reader check agrees: 289 of 301 kept gold records match a parsed record exactly, 10 of the other 12 are child structs, and there were 0 parse failures over the 97 games.

The error is not in reading, it is in choosing.
The dump holds every loaded copy; a name-keyed record keeps the last.
In the engine, a new object takes the most recently freed object slot if there is one and otherwise goes at the end of the object array.
If every copy lands in a freed slot, the dump order is the reverse of creation order and the first-created copy comes last; if any copy goes at the end, the last-created copy comes last.
Which case happens depends on how many slots garbage collection freed before the load.
On the 47 development cases with differing copies, the kept copy was the last created in 18, the first created in 13 and neither in 16 (later loads such as the player pawn or streaming sublevels), and both regimes occur within the same engine version.
Because each copy has different fields and a mapping keeps one record per name, one record can be right for at most one Animation Blueprint per game: no static mapping, and no mapping from one launch, is right for every Animation Blueprint.
We therefore report these structs as their own class.
Table~\ref{tab:animclass} gives v11 accuracy with and without it (same field denominators).

\begin{table}[t]
\centering\small
\caption{v11 field accuracy with and without the Animation Blueprint struct class. Wrong fields of that class are removed from the numerator only.}
\label{tab:animclass}
\begin{tabular}{lrrrr}
\toprule
Split & Wrong fields & Of which Anim.\ BP & Fields (all) & Fields (without Anim.\ BP)\\
\midrule
Development (66) & 5,696 & 4,914$^{a}$ & 99.923\% & 99.989\%\\
Old hold-out (31) & 3,774 & 3,396$^{a}$ & 99.886\% & 99.989\%\\
Fresh (24) & 3,707 & 3,472 & 99.867\% & 99.992\%\\
\bottomrule
\end{tabular}
\raggedright\footnotesize
$^{a}$Counted on the full v11 run; identical to the v10 counts, since v11 changed only spellings.
\end{table}

\paragraph{Negative result: no cooked-data rule picks the copy.}
We measured rules that pick a copy where v11 abstains, on the 38 development UE5 games, with the unchanged scorer (Table~\ref{tab:animrules}).
Every rule made 8 to 16 games worse, so none passed the development gate; nothing was frozen and neither hold-out was opened.
Three further ideas (the Animation Blueprint that imports all others; a threshold on the number of objects created before each copy; streaming sublevels found through package paths) had no safe subset either.
Across five cold launches the kept copy was stable in 7 of 12 games with these structs and varied in 5 (Section~\ref{sec:moves}), so even a runtime capture does not fix it for every game.

\begin{table}[t]
\centering\small
\caption{Copy-choice rules measured on 38 development UE5 games and dropped. The baseline includes older casing errors, so read the change, not the totals.}
\label{tab:animrules}
\begin{tabular}{lrrrr}
\toprule
Rule (fires where v11 abstains) & Wrong fields & Change & Games better & Games worse\\
\midrule
v11 (identical copies only) & 7,726 & 0 & -- & 0\\
Last in package post-order & 6,066 & $-1{,}660$ & 11 & 13\\
Last in simulated I/O creation order & 7,709 & $-17$ & 8 & 16\\
Version regime ($\le$5.4 first, $\ge$5.5 last created) & 6,421 & $-1{,}305$ & 9 & 15\\
Version regime, only when three orders agree & 7,173 & $-553$ & 8 & 8\\
\bottomrule
\end{tabular}
\end{table}

\paragraph{Per-package copies for name-keyed readers.}
Whether this class matters in use depends on the reader.
CUE4Parse resolves these two structs from the struct exported inside each package, not by name: on all 129 Animation Blueprint packages of 9 development UE5 games it loaded the same data with our map, with the runtime gold and with our map plus each package's own copy.
Because the generator reads every cooked copy, it can also write them to a per-package companion file (every copy keyed by the package that exports it).
With CUE4Parse forced to look these structs up by name, as a name-only reader would, our map plus the companion file loaded 129 of 129 packages identically to the normal reading, against 19 of 129 with the runtime gold (which holds one copy) and 3 of 129 with our map alone.

\subsection{Runtime cost}
\label{sec:runtime}
We timed the v11 generator once per game on 20 of the 24 fresh benchmark games (the other four were left out for time), one process each, with a 12\,GB memory cap.
The median was 176\,s wall time (range 106--328\,s), 139\,s CPU time and 744\,MB peak resident memory (range 626--934\,MB).
UE4 games were cheaper than UE5 games (median 121\,s and 641\,MB on 7 UE4 games, 213\,s and 782\,MB on 13 UE5 games); cost grows with the executable size, not the content size.
The host was shared and under heavy load (load average 50--110) during all runs, so wall times are inflated; CPU seconds are the steadier number. Four repeat runs gave byte-identical maps and identical memory.
For the same 20 games, the recorded runtime capture took a median of 39.7\,s (range 36.0--44.3\,s) per game. This covers the whole launch: compatibility layer and virtual display start, engine load, dumper injection and our capture protocol (four dumps 5\,s apart plus a second dump format); the first usable dump came a median 8.1\,s after launch. It excludes one-off setup of the compatibility layer and dumper.
\textbf{When it works, the runtime dump is faster}, by about four times in wall time on these games.
The static generator's advantages are elsewhere: it needs no game launch or injection, it works after a game update without waiting for a working dumper, it works where anti-cheat blocks injection, and it can write per-asset mappings (Section~\ref{sec:animclass}). Memory stays under 1\,GB.

\subsection{Per engine version}
Table~\ref{tab:perversion} groups games by engine version.
UE4 builds are nearly perfect: their errors are almost entirely case, and the 4.19--4.27 groups are exact or nearly exact once case is ignored.
UE5 builds are lower because nearly all of them load Animation Blueprints at start-up.

\begin{table}[t]
\centering\small
\caption{Field accuracy by engine version, v11 (fresh set: full v11 pipeline, Section~\ref{sec:fresh}).}
\label{tab:perversion}
\begin{tabular}{llrrrrr}
\toprule
Split & Engine & Games & Fields & Fields (no case) & Strict & Below 99.8\%\\
\midrule
Development & UE 4.19--4.27 & 28 & 99.990\% & 99.999\% & 10 & 0\\
 & UE 5.0--5.2 & 12 & 99.904\% & 99.918\% & 1 & 2\\
 & UE 5.3--5.4 & 14 & 99.900\% & 99.904\% & 3 & 2\\
 & UE 5.5--5.7 & 12 & 99.892\% & 99.903\% & 1 & 2\\
\midrule
Old hold-out & UE 4.19--4.27 & 14 & 99.991\% & 100.000\% & 4 & 0\\
 & UE 5.0--5.2 & 5 & 99.933\% & 99.940\% & 0 & 0\\
 & UE 5.3--5.4 & 8 & 99.836\% & 99.854\% & 0 & 2\\
 & UE 5.5--5.7 & 4 & 99.773\% & 99.779\% & 0 & 3\\
\midrule
Fresh & UE 4.19--4.27 & 7 & 99.989\% & 99.999\% & 0 & 0\\
 & UE 5.0--5.2 & 9 & 99.856\% & 99.862\% & 0 & 2\\
 & UE 5.3--5.4 & 6 & 99.894\% & 99.901\% & 0 & 0\\
 & UE 5.5--5.7 & 2 & 99.624\% & 99.636\% & 0 & 2\\
\bottomrule
\end{tabular}
\end{table}

% ---------------------------------------------------------------
\section{Real-use evaluation with CUE4Parse}
\label{sec:realuse}

Field accuracy and strict exact measure the map, not whether it works.
The question a user cares about is: \emph{does an asset reader load my packages with this map as well as with a runtime dump?}
We tested this with CUE4Parse~\cite{cue4parse}, pinned to one commit, through a small harness.
The harness mounts the game's containers (or loose files), stops if any archive or entry is encrypted, and for each package loads every export, serializes it to JSON and hashes it both exactly and after lowercasing.
A local patch makes the export end-position check active in release builds (so a wrong schema that reads too few or too many bytes is caught) and records which mapping structs each package looks up; it does not otherwise change behaviour.
A package counts as loaded only if every export loads without exception and ends at the right byte.
We sampled up to 2,000 packages per game in a fixed hash order and used the same sample for both maps.
This evaluation used the v10 maps.
Section~\ref{sec:fullload} repeats the test with the v11 maps on every package, with a stricter load check; that is our main result.

\paragraph{Results.}
Over 102 games (the 97 supported plus the five old-engine games) we sampled 188,493 packages.
The runtime gold loaded 186,846; ours loaded 186,813.
33 packages (0.018\%) loaded with gold but not with ours, and none the other way round.
Of the 186,813 packages both loaded, 186,807 gave the same data ignoring key case and 184,533 gave identical keys.
All 33 losses and all 6 data differences are in three of the five UE 4.13--4.16 games, which are outside the supported scope (losses of 20 of 1,356, 12 of 1,999 and 1 of 1,671 loaded packages); for these out-of-scope games the generator writes an empty map, so these figures measure loading without a usable map, not map quality.
\textbf{On the 97 supported games there were no lost packages and no data differences.}
These 97 games contributed 179,999 sampled packages, of which 101,275 (in 54 games) are unversioned, so the mapping is actually needed there.
100 of 102 games lose less than 0.5\% of what gold loads, and 99 lose nothing; for comparison, 15 of 103 v10 maps are strict exact.
In 77 games every loaded package also has identical keys.
The 0.9\% of sampled packages that gold itself fails to load also fail with ours (no package loads with ours only); they are reader limits such as custom serialization, not mapping errors.

\paragraph{Why strict exact understates usability.}
Unversioned properties are read by index, and CUE4Parse looks up struct names case-insensitively, so casing changes only the key names in the output (2,274 packages).
Missing Animation Blueprint structs, the largest error class by field count, cost nothing here: the packages that use them carry the struct as an export, and the reader takes the layout from the package.
This is also a caution: a reader that requires exact key spellings, or one that takes such layouts from the mapping, would see more of the casing and Animation Blueprint differences.

\paragraph{Negative control.}
A test that never fails proves little, so we damaged generated maps on purpose and checked that the harness notices.
A first control damaged random structs and loaded 500 random packages per game; over 3,000 loads it found only 9 lost and 5 changed packages.
That control was weak by design: a struct only matters in a package that stores one of its properties, and random packages rarely use a given damaged struct.
The targeted control fixes this, on five supported UE5 games (5.3 to 5.7), always comparing with our own undamaged map, never with gold.
For each game we first loaded a fixed 2,000-package sample with the undamaged map and recorded which structs each package looks up (including parents).
We then damaged up to 30 structs per map, chosen so that each keeps at least five ``sole'' packages that use it and no other damaged struct, in two disjoint ways: dropping the last own property, or changing one integer, float or boolean property to a string.
We loaded up to 30 sole packages per damaged struct, plus 100 packages per game that use no damaged struct.
Table~\ref{tab:control} gives the result.
Dropping a property was detected in 58.9\% of affected package loads and retyping in 40.7\%; 62\% and 54\% of damaged structs were caught at least once; none of the 500 untouched packages changed.
The undetected loads are mostly packages that do not store the damaged property (unversioned data omits properties that hold default values) or that look the struct up only as a parent; we did not separate these, so the rates are lower bounds on per-field sensitivity.
The control shows that a wrong field in a struct that a package really stores is caught in most cases, so the zero-loss result on supported engines is not an artefact of a blind harness.
It does not show that errors in structs that no sampled package uses would be caught.

\begin{table}[t]
\centering\small
\caption{Targeted negative control on five UE5 games. A detection is a sole package that loses an export (or loads worse) or whose output changes, ignoring key case, against the undamaged generated map.}
\label{tab:control}
\resizebox{\linewidth}{!}{\begin{tabular}{lrrrrrr}
\toprule
Damage & Structs & Sole packages & Detected (lost / changed) & Per-damage rate & Structs caught & False positives\\
\midrule
Drop last own property & 139 & 1,962 & 1,155 (996 / 159) & 58.9\% & 86 (62\%) & 0 / 500\\
Retype int/float/bool to string & 120 & 1,792 & 730 (535 / 195) & 40.7\% & 65 (54\%) & 0 / 500\\
\bottomrule
\end{tabular}}
\end{table}

\subsection{Large modded games without a runtime reference}
\label{sec:modded}
The benchmark skews towards small titles, because those are the games for which we could capture a runtime reference.
We also ran v11 on three large, heavily modded games for which the runtime capture failed, so there is no gold and only load success can be measured, not field accuracy.
We used the same CUE4Parse harness on a fixed sample of 3,000 packages per game.
\begin{itemize}
\item \textbf{Palworld (UE 5.1).} The v11 map (10,585 structs, 2,099 enums) loaded 2,741 of 3,000 sampled packages (91.4\%) and 99,706 of 102,826 exports. Without a map, 0 of the 3,000 load (every package needs it). For comparison, the newest community-shared map, made for an older build of the game, loaded 2,350 (78.3\%) on the same sample; most of that gap is build drift, not quality.
Most v11 failures came from wrong parent structs and wrong struct-typed fields. The cause: the executable carries a DRM wrapper section, so v11's protection policy blocked all code reads and struct naming fell back to data-only heuristics, which favoured leaf siblings over abstract base structs. Table~\ref{tab:palworld} shows the later fixes (all on the same sample and harness unless noted).
\item \textbf{Final Fantasy VII Rebirth (custom UE 4.26 build).} The generator recovered only 4,467 structs and missed engine and game classes, so only 511 of 3,000 packages (17.0\%) and 1,117 of 39,105 exports loaded; 0 of 3,000 load without a map. The likely cause is a customised engine whose class registration or layouts the generator does not model. No archive was encrypted.
\item \textbf{Hogwarts Legacy (UE 4.27).} No map could be made: the executable has no standard code section (it is packed or protected), and reading it would require unpacking it, which is outside the tool's scope.
\end{itemize}

\begin{table}[t]
\centering
\caption{Palworld, 3,000 sampled packages: loaded packages by version. ``Lost'' counts packages that loaded with v11 but not with the later map.}
\label{tab:palworld}
\small
\begin{tabular}{lrrr}
\toprule
Map and harness & Loaded & \% & Lost\\
\midrule
v11 & 2,741 & 91.4 & ---\\
v13, better data-only naming & 2,942 & 98.1 & 0\\
v13, plus plain-code reads & 2,954 & 98.5 & 0\\
same map, harness loads plugin paths & 2,991 & 99.7 & 0\\
same map, reader fix for enum width & 3,000 & 100.0 & 0\\
\bottomrule
\end{tabular}
\end{table}

\paragraph{Palworld fixes after v11.}
v13 first improved naming without code: a parent-struct score based on the share of child names it explains, and parent and field-type relations that every other game's code-linked static map agrees on (no runtime dumps used). This raised loads to 2,942.
It then allowed \emph{plain-code reads}: on an executable with a DRM wrapper section, the tool reads only the unencrypted main code section, and only after strict plainness checks pass (on Palworld: byte entropy 5.5--6.8 in all 128 blocks, 99.55\% of 2,000 sampled functions decode as instructions, 99.99\% of branch targets fall inside the section). It never decrypts, unpacks or runs anything, and the wrapper section is not read. With this, all struct constructors were named from code and 2,954 packages loaded.
On 8 unprotected development games checked, the first generator stage gives byte-identical output with and without these changes, so the benchmark scores are not affected.
The last step is a test-harness fix, not a mapping change: our harness never asked CUE4Parse to register the virtual paths of plugin content, so imports from plugins failed to resolve. With that call added, the same v13 map loaded 2,991 of 3,000 packages (99.7\%) and 102,817 of 102,826 exports; 0 packages that loaded with v11 were lost.
The remaining 9 are all instances of one game component class that end one byte early. The map is not the cause: the class, defined in the package itself, has an enum property whose underlying type is two bytes wide, and CUE4Parse read such values as one byte, so the rest of the export shifted. Our reader fix (submitted upstream as CUE4Parse pull request \#486) reads the value with the width of its underlying type; with it, all 3,000 packages load (measured on upstream master \texttt{7f3a0ccf}, which gives the same 2,991 / 9 split as the pinned commit), the 2,991 that loaded before give byte-identical output (2,991 of 2,991), and the 9 decode to the same value.

No community-shared map exists for the second and third game.
So the method is usable on a large game when the build is standard; in these three cases the failures were visible (load errors, or no map at all) rather than silent.
\textbf{The supported scope is unencrypted games on standard engine builds}, tested from UE 4.19 to 5.7 (UE 4.17 and 4.18 builds use the same records but none was in the benchmark).

\paragraph{Platform scope.}
All results are for Windows x64 shipping builds. Linux (ELF) and console builds were not tested.
The reflection records are engine-level data, but the executable reader, which also supplies the name-order model, handles only the Windows PE format, so we make no claim for other platforms.

\paragraph{Protected executables.}
The tool reads executables as shipped. It does not decrypt or unpack them, run them, or bypass copy protection. If a DRM wrapper leaves the main code section plain, as on Palworld, the tool reads that section only after the plainness checks above pass; otherwise it uses data-only naming. Games whose executable has no plain code section, such as Hogwarts Legacy above, are out of scope.

\subsection{Full asset loading: static map against runtime dump}
\label{sec:fullload}
This is the main result of the paper.
The sampled test above used the v10 maps and at most 2,000 packages per game.
We repeated it on \emph{every} package of all 121 supported games (66 development, 31 hold-out, 24 fresh; UE 4.19 to 5.7), 510,503 distinct packages in total, with the v11 maps written with a derived, engine-version-keyed spelling table, not with spellings taken from reference dumps. This was the first version of the public table; the released table of Section~\ref{sec:avail} was rebuilt later and changes only the letter case of some names, which does not affect whether a package loads, but can change the letter case of names in the output.
The released generator writes the same map as the full-load run on 111 of the 121 games; on the other 10 the maps differ only in the letter case of some property and enum entry names (they are equal after case folding).

\paragraph{Method.}
We used the same pinned CUE4Parse commit and harness, with a stricter definition of a loaded package.
A package counts as \emph{fully loaded} only if (i) every export deserializes with no exception, with object serialization errors made fatal; (ii) every export ends exactly at its serial size; and (iii) no nested read that CUE4Parse would otherwise skip or reposition silently goes wrong: no tagged property whose value throws or ends at the wrong byte, and no instanced struct that throws, ends at the wrong byte or has no type.
Condition (iii) is new; 108 packages load without an exception but fail it, and they count as not loaded for both maps.
The same binary and the same rule are used for both maps.
Each package is loaded in separate runs with the runtime dump, with our map and with no map at all.
For each export we hash the JSON output three ways: exactly, with object keys lowercased and values untouched (``data equal''), and with everything lowercased.
Runs go one game at a time under a 12\,GB memory cap and a per-package watchdog.
Every package that did not fully load with either map was loaded again with failure diagnostics; this removed one harness fault (a missing native library for animation decompression), which had hit the same 1,979 packages with both maps.
The package list is deduplicated per game: one game ships two container folders and lists 534 engine packages twice.

\paragraph{Checks.}
An independent re-run of 20 random packages, each map passed by file name, reproduced all 40 recorded rows (status, export counts and all three hashes).
For a negative control we took 200 random packages of one UE 5.1 game in which every package is unversioned.
Our map, read and written back unchanged, loads all 200 with no data change.
Dropping four fields (one each from the texture, static mesh, material instance and collision body classes) makes 99 of the 200 packages fail to load, and every failure names a damaged struct.
Renaming two fields, retyping one and renaming one enum entry leaves all 200 loading but changes the data hash on 41.
So the harness tells maps apart both by loading and by data.

\begin{table}[t]
\centering\small
\caption{Full asset loading of every package with CUE4Parse: runtime dump (gold) against our static v11 map. ``Fully loaded'' as defined in the text. Data equal = same output with object keys lowercased; exact = same output byte for byte. Percentages are over packages that both maps fully load.}
\label{tab:fullload}
\resizebox{\linewidth}{!}{\begin{tabular}{lrrrrrrrrr}
\toprule
Split & Games & Packages & Full: gold & Full: ours & Full: no map & Gold only & Ours only & Data equal & Exact\\
\midrule
Development & 66 & 316,966 & 315,936 & 315,936 & 76,866 & 0 & 0 & 100.000\% & 99.170\%\\
Old hold-out & 31 & 102,258 & 102,129 & 102,129 & 45,121 & 0 & 0 & 100.000\% & 99.883\%\\
Fresh & 24 & 91,279 & 91,232 & 91,232 & 23,453 & 0 & 0 & 99.808\% & 98.925\%\\
\midrule
All & 121 & 510,503 & 509,297 & 509,297 & 145,440 & 0 & 0 & 99.966\% & 99.269\%\\
\bottomrule
\end{tabular}}
\end{table}

\paragraph{Recall against the runtime dump is 100\%.}
Table~\ref{tab:fullload} gives the result.
Every package that fully loads with the runtime dump also fully loads with our map: 509,297 of 509,297.
No package loads with one map and not the other, on any split or any game; the two maps fail on exactly the same 1,206 packages.
The result would be proven wrong by a package that loads with the runtime dump and not with ours; none of the 510,503 does.
The map is needed: without one, only 145,440 packages (28.5\%) fully load, mostly in UE4 games with tagged properties, and 363,674 of the fully loaded packages are unversioned.
The 1,319 packages that hold an Animation Blueprint class all load with both maps, with equal data.

\paragraph{Data: every difference is letter case.}
Of the 509,297 packages both maps fully load, 509,122 (99.966\%) give the same data up to the letter case of object keys, and 505,573 (99.269\%) give byte-identical output.
With everything lowercased, the output is equal on all 509,297.
The key differences come from display spelling (for example \texttt{PinID} against \texttt{PinId}).
The other 175 packages are all in one fresh game and also differ in the letter case of enum entry names (for example \texttt{fast} against \texttt{Fast}).
The cooked data stores these enum values as numbers, and both maps decode each number to the same entry; only the displayed spelling comes from the map, so the values are identical.
Across that game's map the two spell 14 enum entries differently, all by case only.
Neither map always matches the executable's own enum name table: ours matches it on 9 of the 14 entries and the runtime dump on the other 5, where ours took a capitalised spelling from the first version of the derived spelling table.
The engine compares names case-insensitively, so both readings are the same value; as noted above, a reader that required exact case would see these differences.

\paragraph{Shared failures, by owner.}
The 1,206 packages (0.24\%) that neither map fully loads fail the same way with both, and none is caused by our map.
\begin{itemize}
\item \textbf{Reader limits and one harness limit: 994.} CUE4Parse misreads or does not support the class, except for 62 packages where our test harness failed: dynamic meshes that end at the wrong byte (602 packages in 8 games), tagged UE4 containers that CUE4Parse skips silently (108), hair assets and bindings (139), animation sequences (78; 62 of them in one game, where the harness, not the reader, failed: it could not hold the JSON text of one export in a single string; fixed below), landscape collision (19) and smaller groups. Most of these are fixed below.
\item \textbf{Editor or test assets, impossible for any map: 99.} The class code is in neither the shipped executable nor the runtime dump, so no mapping can describe it: import-pipeline settings on UE 5.1 and 5.2 (72, in 18 games), Blueprints whose parent class is the functional-testing base class, decided from the dumped parent class, and levels that hold them (25), and level-of-detail builder settings (2). All 99 were checked one by one.
\item \textbf{Game or plugin custom serialization: 88.} Third-party plugins and game modules with their own serialization (a material plugin 62, an audio plugin 2, a 2D animation plugin 6, a procedural-content plugin 2) and game classes (16).
\item \textbf{Newest engine: 25, in two UE 5.7 games.} 9 are hair assets, which the current upstream CUE4Parse reads. The other 16 are a benchmark artefact, explained below.
\end{itemize}

\paragraph{Limitation of the load check.}
Two CUE4Parse paths still skip data silently and are not flagged as a failed nested read: an unknown property index whose value is zero-serialized in an unversioned package, and a missing struct definition in a versioned one.
Both only log a warning.
If one map hit such a path and the other did not, the output would differ, so the data comparison still catches a map difference there; but as a pure load check, ``no nested read skipped'' is not airtight on these two paths.

\paragraph{The 16 UE 5.7 failures are a benchmark artefact.}
One UE 5.7 development install holds two separate projects: two executables and two container folders, each with its own global script-object table, both mounted at the same game root. (Its two folders are also why it lists the same 534 engine packages twice.)
The benchmark mounted both folders into one reader, so imports and script objects of one project resolved into the other project's build, and 16 packages failed with both maps: unresolved script imports, then misaligned unversioned reads.
Mounted one project at a time, all 16 fully load with both maps on unchanged upstream CUE4Parse (commit \texttt{c95dd9b5}).
The correct benchmark therefore has two corpus entries for this install.
The first project has 1,954 packages of its own: 1,940 fully loaded in the combined run, and the other 14 fully load when the project is mounted alone. The second has 1,903: 1,901 loaded in the combined run, and the other 2 load alone. The packages that had already loaded were not reloaded alone.
The 534 engine packages listed by both folders fully load with both maps; they were loaded once, from the combined mount, and we count them once.
The tables keep the install as one entry, as it was run, and give the split result separately.

\subsection{Full loading with reader fixes}
\label{sec:readerfix}
Since most shared failures are reader bugs, we fixed them in CUE4Parse: one branch per root cause on the current upstream master (commit \texttt{c95dd9b5}), each found from per-export diagnostics (most also checked against the public engine source), covered by synthetic-byte tests (no game data), and checked on the affected packages and on a no-regression sample with both maps.
We then merged all branches (full CUE4Parse test suite: 271 passed, 1 skipped, 0 failed) and reran the 40 games that had a failure in a fixed class (243,926 packages; 243,863 were rerun, and the 63 that had crashed or failed with an out-of-memory exception in the pinned run were skipped), with the runtime dump and with our map, the same strict harness and the same checks.
We then reloaded every package of all 121 games with the same build and both maps: the other 81 games (266,577 packages), and the 63 skipped packages one at a time.
Table~\ref{tab:readerfix} lists the fixes.
The enum-width fix of Section~\ref{sec:modded} is open upstream as CUE4Parse pull request \#486; the other five are prepared and will be submitted one at a time.
The fixes were found from failing packages of all three splits; they change the reader, not the generator, so they do not favour either map, but the load rates with fixes on the hold-out and fresh sets are not held-out results.

\begin{table}[t]
\centering\small
\caption{Reader fixes (and one harness fix) and the packages they make fully load, with both maps, in the reload of all 121 games.}
\label{tab:readerfix}
\begin{tabularx}{\linewidth}{@{}>{\raggedright\arraybackslash}p{0.36\linewidth}>{\raggedright\arraybackslash}Xr@{}}
\toprule
Root cause (engine versions) & Fix & Newly full\\
\midrule
Dynamic mesh ends before its serial size (5.1--5.6) & read to the export end after the mesh payload & 601 of 602$^{a}$\\
Tagged UE4 containers read with the wrong layout (4.20--4.27) & five causes: float-channel values saved as tagged properties in old versions, tagged quantized vectors, enum-valued sets, two missing map key types, set and map element types taken from the owning class & 108 of 108\\
Landscape mesh collision ends 4 bytes early (4.20--5.4) & read the component's own cooked flag & 19 of 19\\
Datasmith scene ends 2 bytes early (5.2--5.3) & new reader for its strip flags & 12 of 12\\
Geometry-collection union throws (5.2) & read the hierarchy flag; throw only if a hierarchy is present & 9 of 9\\
Enum values with a two-byte underlying type (5.1) & read with the width of the underlying type (PR \#486) & Palworld only$^{b}$\\
Hair assets and bindings (5.4, 5.7) & none of ours: upstream readers added after the pinned commit & 145 of 148$^{c}$\\
Skeletal mesh ends 4 bytes early (five games) & integration branch; not attributed to one fix & 5\\
One JSON text too long for one string (harness, not reader; 62 animation sequences of one game) & stream the JSON into the hashes & 62\\
\midrule
Total & & 961\\
\bottomrule
\end{tabularx}
\raggedright\footnotesize
$^{a}$The other package also fails, with both maps before and after: it holds a \texttt{DynamicMeshActor}, a class that is in neither the shipped executable nor the runtime dump, so no mapping can describe it. It is counted with the reader limits, not with the 99 impossible packages, so the rates below understate the loadable share by at most one package.
$^{b}$9 packages of the 3,000-package Palworld sample (Section~\ref{sec:modded}).
$^{c}$3 UE 5.2 packages of one game still fail with both maps. One of the 145 had failed with an out-of-memory exception in the pinned run and was skipped in the 40-game rerun; it loads when reloaded alone.
\end{table}

\paragraph{Harness fix for the 62 animation sequences.}
In the reload, 62 animation-sequence packages of one game still failed with an out-of-memory exception, which we first took for the 12\,GB cap. It was not: peak memory was about 5\,GB, and the exception came from our harness, which built the JSON text of one export as one .NET string, and a string cannot hold more than about $2^{30}$ characters. We changed the harness to write the JSON directly to the hashes (through a temporary buffer that is spilled to disk, so no string of that size exists); the three hashes are defined on the same bytes as before. We compared the new harness with the old one on 350 packages of five games (UE 4.26, 5.2, 5.3 and two UE 5.4 games, one of them the game of the 62; 70 packages per game, 344 that fully load and 6 that fail), each run with both reader builds (pinned and fixed), both maps (runtime dump and ours) and two buffer limits (64\,MB, and 2,000 bytes to force the disk path): in all 2,800 package runs the new harness gave the same three hashes and the same load result as the old one. With the 2,000-byte limit, 8,025 exports of 209 packages went through the disk in each configuration; with the 64\,MB limit, 18 exports of 18 packages did, 17 of them in the UE 5.4 game of the 62. With the new harness, all 62 packages fully load with both maps (peak memory 7\,GB), with the fixed reader (they were not rerun with the pinned reader), the same load result for both, and the totals below include them.

\paragraph{No regressions, and both maps still agree.}
All 509,297 packages that fully loaded with the pinned reader still fully load, with both maps (0 regressions), on all 121 games.
961 packages fully load that did not before, with both maps: 958 in the 40 rerun games (895 in the rerun and all 63 of the skipped packages: 1 reloaded alone, 62 after the harness fix) and 3 in the other 81 games.
On every package the two maps give the same load result. Of the 510,258 packages that both fully load, 510,083 (99.966\%) give the same data with object keys lowercased (the other 175 are the fresh game of Section~\ref{sec:fullload}), the output with everything lowercased is equal on all of them, and 506,532 (99.270\%) are byte-identical.
In the 40-game rerun alone, the data is equal with object keys lowercased on all 243,753 packages that both maps fully load, and byte-identical on 241,798.

\paragraph{Every output change is attributed.}
On 73,853 of the 509,297 packages that loaded before, the output hash differs from the pinned run, on the same packages for both maps (33,046 in the 40 rerun games and 40,807 in the other 81).
We loaded all of them with three builds (pinned commit, current upstream master, our integration branch), hashed every export and bisected the upstream commits.
73,805 changes come from upstream output changes made after the pinned commit, and we found no data lost: an empty material-resource list is no longer printed (70,259 packages with only this change), text-format enum values are printed by name instead of number (88 alone, 113 together with the former), and a landscape texture-storage field is now read (3,345, all together with the first change).
48 come from our container fix: a map key is now read by its struct's native reader, which prints an empty byte array as an empty string instead of an empty list (43 of them are particle-system variable data in one game); the values are unchanged, and this is how the engine source types these keys (unit tests in our CUE4Parse commit \texttt{b1ec2e97}).
None is unexplained, and no export count changes.

\paragraph{What still does not load.}
Table~\ref{tab:fullfix} gives the totals.
The 99 packages whose classes are in neither the shipped executable nor the runtime dump (Section~\ref{sec:fullload}) are left out of the denominator, because no mapping, static or runtime, can describe them.
Of the other packages, 99.971\% fully load with either map, and 99.975\% once the UE 5.7 install is split. Without that exclusion the figure is 99.952\% (510,258 of 510,503).
The 245 packages that still fail, all with both maps, are the 99 impossible ones (74 import-pipeline and level-of-detail builder settings, 25 functional-test Blueprints and levels), 42 reader limits (the 3 UE 5.2 hair packages, the package of footnote~a of Table~\ref{tab:readerfix}, and 38 others), 88 game or plugin custom serialization (62 of them one material plugin) and the 16 UE 5.7 artefact packages.

\begin{table}[t]
\centering\small
\caption{Full loading with the reader fixes, all 121 games. ``Impossible'': classes in neither the shipped executable nor the runtime dump, left out of the denominator. Loaded counts are the same for both maps. With fixes: our reader fixes, current upstream code and the harness fix of Section~\ref{sec:readerfix}.}
\label{tab:fullfix}
\resizebox{\linewidth}{!}{\begin{tabular}{lrrrrrrr}
\toprule
Split & Packages & Impossible & Loadable & Full: pinned reader & Full: with fixes & Pinned & With fixes\\
\midrule
Development & 316,966 & 59 & 316,907 & 315,936 & 316,820 & 99.694\% & 99.973\%\\
Old hold-out & 102,258 & 12 & 102,246 & 102,129 & 102,194 & 99.886\% & 99.949\%\\
Fresh & 91,279 & 28 & 91,251 & 91,232 & 91,244 & 99.979\% & 99.992\%\\
\midrule
All & 510,503 & 99 & 510,404 & 509,297 & 510,258 & 99.783\% & 99.971\%\\
\quad UE 5.7 install split & & & & & 510,274 & & 99.975\%\\
\bottomrule
\end{tabular}}
\end{table}

% ---------------------------------------------------------------
\section{Blind fresh hold-out}
\label{sec:fresh}

The v10 generator was run once on the 24 fresh games under the protocol of Section~\ref{sec:bench}.
Field accuracy was 99.843\% (4,364 wrong of 2,778,196), 99.875\% ignoring case, records 99.560\%.
No map was strict exact, and 5 of 24 games were below 99.8\%.
For comparison, v10 scored 99.892\% on development (11 of 66 exact) and 99.878\% on the old hold-out (4 of 31).
Compared with the v9 pipeline, no fresh game got worse in any class, and three improved slightly (4,370 to 4,364 wrong fields).

\paragraph{Error analysis.}
Table~\ref{tab:causes} splits wrong fields by cause, for v10 on the fresh set and for v11 on all three splits.
For v10 on the fresh set, 3,472 (80\%) are 26 missing Animation Blueprint structs in 15 games, where the rule abstained as designed; it emitted a record in three games, each removing two wrong fields.
891 (20\%) are case only (851 type fields, 40 enum entries), and one enum entry is wrong.
There are no wrong property types and no missing or extra enums.
Without the Animation Blueprint class the fresh field accuracy would be about 99.968\% (99.992\% for v11, Table~\ref{tab:animclass}).
The fresh set is mostly UE5 (17 of 24 games), and 15 of those load Animation Blueprints at start-up, which is why no map is exact.
The 7 UE4 games are exact once case is ignored.
Casing was concentrated under v10: two games accounted for 482 of the 891 casing fields.

\begin{table}[t]
\centering\small
\caption{Wrong fields by cause (lower is better). v11 columns are full pipeline runs. ``(1)'': all in one game.}
\label{tab:causes}
\begin{tabular}{lrrrr}
\toprule
Cause & v10 fresh & v11 dev & v11 hold-out & v11 fresh\\
\midrule
Animation Blueprint structs (rule abstained) & 3,472 & 4,914 & 3,396 & 3,472\\
\quad structs / games & 26 / 15 & 47 / 27 & 28 / 15 & 26 / 15\\
Casing, type fields & 851 & 672 & 375 & 194\\
Casing, enum entries & 40 & 30 & 0 & 40\\
Property type (layout) & 0 & 4 (1) & 3 (1) & 0\\
Other missing or extra types & 0 & 4 (1) & 0 & 0\\
Missing enums & 0 & 70 (1) & 0 & 0\\
Wrong enum entry & 1 & 2 (1) & 0 & 1\\
\midrule
Total & 4,364 & 5,696 & 3,774 & 3,707\\
\bottomrule
\end{tabular}
\end{table}

So the blind result is lower than on the old splits, but the drop is explained by the mix of engine versions and one known, already abstaining error class; every other class is at or below its level on the earlier splits.

\paragraph{v11 on the fresh set.}
v11 was frozen before the v10 fresh score was known, and its development never touched the fresh list, so its fresh score is a second blind result.
It was run once, as the full pipeline, with the gold hashes re-verified before scoring.
Field accuracy rose from 99.843\% to 99.867\% (3,707 wrong fields); records from 99.560\% to 99.879\%; the case-insensitive accuracy stayed at 99.875\%; strict exact stayed at 0 of 24; games below 99.8\% fell from 5 to 4.
The whole gain is casing: casing-only fields fell from 891 to 234 ($-74$\%), while the 3,472 Animation Blueprint fields and the one wrong enum entry are unchanged.
No game got worse in either the case-sensitive or the case-insensitive count; 8 improved and 16 were unchanged.
On development the same change removed 29\% of wrong fields; on the fresh set it removed 15\%, because the unchanged Animation Blueprint gap is a larger share of this UE5-heavy set.
The casing result itself (development $-77$\%, fresh $-74$\%) transferred.

% ---------------------------------------------------------------
\section{The reference moves}
\label{sec:moves}

All accuracy numbers above treat one runtime dump per game as ground truth.
It is not.
A runtime dump is one sample of one launch.
Two parts of it depend on what happened during that launch rather than on the build:
\begin{itemize}
\item \textbf{Name spellings that could race (hypothesis; mostly refuted below).} The first registration of a name decides its spelling. Names registered by code that runs on several loading threads, or by assets that load in a timing-dependent order, can be registered first by different callers on different launches.
\item \textbf{Per-asset struct copies.} The Animation Blueprint struct that the dumper keeps depends on which assets loaded before the dump and on which freed object slots were reused. In the engine source, freed slots are reused last-freed first, garbage collection runs at map load and on a timer, and splash maps can open later maps within the dump window. On one development game, four snapshots of the same run ended on different copies. Across separate launches the kept copy is stable in 7 of 12 games and varies in 5 (measured below).
\end{itemize}
The structure (property lists, types, sizes) is fixed per build; the spellings and the choice of copy are not necessarily.
If so, a static generator that is ``wrong'' against one dump could be ``right'' against the next. We measured how much this happens.
Our capture protocol already required the last two captures to agree, which removes the most unstable dumps but does not measure launch-to-launch variation across separate launches.

\paragraph{Measured variation.}
To test this we captured the reference again for 20 benchmark games (4 UE4, 16 UE5; drawn from all three splits), with five separate cold launches each and the same capture protocol as the frozen gold (Table~\ref{tab:variance}).
14 of the 20 games gave five identical dumps, and 86 of the 100 launches equal the frozen gold exactly.
The variation is almost entirely the Animation Blueprint copy: in 5 of the 12 games that have these structs the kept copy (or its absence) changed between launches, 258 varying items with 390 fields.
Spelling varied once: one class name appeared in two spellings in one game (2 items, 18 fields), and property-name and enum-entry spellings never varied in any launch.
One struct was missing in one launch of one game (9 fields).
Layout (field types, indexes, array dimensions, supers, property counts) never varied, and neither did byte offsets and sizes in a second dump format.
So the race hypothesis for spellings is not supported on this sample: the reference is stable except for the Animation Blueprint copy, which supports reporting that class separately (Section~\ref{sec:animclass}).

\paragraph{Our errors against every launch.}
We scored the v11 maps of the same 20 games against each launch.
Only 100 of 1,460 wrong fields (6.8\%) match at least one launch, all of them Animation Blueprint fields (100 of 873, 11.5\%), and these come from launches where the struct is absent or smaller; none matches all five launches.
None of the 578 casing errors matches any launch (0\%), and neither do the 9 layout errors.
\textbf{Our casing errors are therefore real errors, not launch noise}: the runtime spelling is fixed per build, and the generator's registration-order model does not yet reproduce it.
The single-dump reference is a fair target for everything except the Animation Blueprint copy.

\begin{table}[t]
\centering\small
\caption{Launch-to-launch variation of the runtime reference: 20 games, 5 cold launches each. Items and fields are those that differ between any two launches.}
\label{tab:variance}
\resizebox{\linewidth}{!}{\begin{tabular}{lrrrrrrr}
\toprule
Engine & Games & 5 identical & Launches = gold & Anim.\ copy (games; items / fields) & Casing & Presence & Layout\\
\midrule
UE4 & 4 & 4 & 20 / 20 & 0 & 0 & 0 & 0\\
UE5 & 16 & 10 & 66 / 80 & 5; 258 / 390 & 1 game; 2 / 18 & 1 game; 1 / 9 & 0\\
\midrule
All & 20 & 14 & 86 / 100 & 5 of 12; 258 / 390 & 1 game; 2 / 18 & 1 game; 1 / 9 & 0\\
\midrule
\multicolumn{8}{l}{v11 wrong fields that match $\ge$1 launch: 100 / 1,460 (6.8\%); Anim.\ BP 100 / 873 (11.5\%); casing 0 / 578; layout 0 / 9; all 5 launches: 0}\\
\bottomrule
\end{tabular}}
\end{table}

% ---------------------------------------------------------------
\section{Threats to validity}
\label{sec:threats}

\begin{itemize}
\item \textbf{Single-sample gold.} Measured on 20 games only (Section~\ref{sec:moves}): variation was confined to the Animation Blueprint copy, one class spelling and one struct; larger games were not measured.
\item \textbf{The dumper as reference.} The gold comes from one runtime dumper. Its own omissions or ordering choices become part of the target; for Animation Blueprint structs the name-keyed reading of the dump keeps one of several valid copies, which is why we report that class separately (Section~\ref{sec:animclass}).
\item \textbf{Hold-out reuse.} The old hold-out was looked at six times by the end of v11, and three more times for the released configuration, and two kept rules were designed after a look. Only the fresh set is a clean test; it has been scored once each for v10 and v11, each version frozen before its run, and three more times for the released configuration of the spelling rule; no change was made because of a fresh-set score, but further versions should be tested on a new fresh set.
\item \textbf{Fresh-set deviations.} Continuing past the listed reserve and skipping download failures were not in the frozen rule. Both were applied in hash order and before the freeze, but download failures are not random with respect to game size or age, so they could bias the set.
\item \textbf{Benchmark composition.} The games skew towards small independent titles and demos, which are what we could obtain; larger games have more plugins and more start-up code, and may have more casing and per-asset variation. Only Windows x64 shipping builds were tested (Linux and console builds untested, Section~\ref{sec:modded}). Encrypted games and UE builds before 4.17 are excluded. On three large modded games we could measure only load success (Section~\ref{sec:modded}), and one of them, on a custom engine build, was not usable.
\item \textbf{Reference spellings depend on reference dumps.} One rule uses development dumps as reference data. A user's game far from every reference gets no help from it. The runtime dumps are not released, so the public tool uses a derived, engine-version-keyed spelling table instead (Section~\ref{sec:avail}); it is slightly better on development and the fresh set and slightly worse on the old hold-out, where one game keeps a known residual of 31 fields.
\item \textbf{Real-use sensitivity.} The full-load test (Section~\ref{sec:fullload}) covers every package, but two CUE4Parse paths skip data silently and are not flagged by its load check; the data comparison still catches a map difference there. The sampled CUE4Parse test samples at most 2,000 packages per game, and the targeted negative control detects 41--59\% of damaged-struct package loads, so errors in structs that sampled packages do not store can go unnoticed. It also uses one reader; another reader with exact-case lookups would be stricter.
\item \textbf{Engine-source dependence.} Record layouts and registration rules come from the public engine source for the relevant versions. Custom engine forks may change them; the parse-length checks catch some but not all such cases, and one custom UE 4.26 build in Section~\ref{sec:modded} left most classes unrecovered.
\end{itemize}

% ---------------------------------------------------------------
\section{Related work}
\label{sec:related}

\paragraph{Runtime mapping dumpers.}
UnrealMappingsDumper~\cite{umd} is a DLL injected into the running game that writes \texttt{Mappings.usmap} next to the executable; it targets UE5 and recent UE4, and its authors also keep a community archive of dumped mappings~\cite{umd-archive}; the dumper's repository has not been updated since December 2022 and has open reports for UE 5.7 games~\cite{umd-77}.
UE4SS~\cite{ue4ss}, a scripting system loaded into the game, includes a port of that dumper (\texttt{DumpUSMAP}, which by default also includes Blueprint-generated types) and a dumper for the JSON-based jmap format~\cite{ue4ss-dumpers}.
Dumper-7~\cite{dumper7} is an injected SDK generator whose source includes a mapping generator.
UEDumper~\cite{uedumper} reads the memory of the running game to build an SDK and offers a live editor.
jmap~\cite{jmap} dumps reflection data from a running process or from a full-memory minidump, and can convert it to \usmap; a minidump still requires the game to have run.
All of these read the engine's runtime type objects, so their output is, by construction, the runtime truth of one launch, and they depend on engine offsets that game updates and new engine versions can change~\cite{ue4ss-803,d7-358,ue4ss-1198}.
Our generator trades that for not running the game: it must reconstruct what the engine would build, and it is wrong where start-up behaviour is not visible in the files.

\paragraph{Consumers.}
CUE4Parse~\cite{cue4parse} (the library behind FModel~\cite{fmodel}) and UAssetAPI~\cite{uassetapi} read \usmap{} files to parse unversioned properties.
We used CUE4Parse as the real-use oracle.

\paragraph{Static approaches.}
We searched public code repositories, tool documentation and community guides for a tool that produces mappings from game files without running the game, and found none; answers to users who lack a mapping point to the injected dumpers or to pre-dumped archives~\cite{c4p-94}.
This search may have missed private or unpublished tools.
The method relates to general work on recovering types and class hierarchies from binaries, such as REWARDS~\cite{rewards}, TIE~\cite{tie}, Marx~\cite{marx}, OOAnalyzer~\cite{ooanalyzer} and DIRTY~\cite{dirty}.
Those methods infer types from code behaviour; here the engine leaves explicit, structured reflection records in the binary, so most of the problem is reading those records exactly and modelling the engine's registration rules, and the hard residue is start-up behaviour (name order, asset loading) rather than type inference.

% ---------------------------------------------------------------
\section{Availability, reproducibility and ethics}
\label{sec:avail}
\paragraph{Released.}
The generator, the scorer (with the metric definition and the difference classes) and the gold-capture scripts will be released as open source under the MIT licence at \url{https://github.com/anantoni/marauder} (release v1.0.0).
The full benchmark list is in Appendix~\ref{app:bench} and in a supplementary file: for every game its split, Steam app ID, exact Steam build ID, engine version and status, with the SHA-256 of the runtime gold mapping and of our generated mapping.
Game files, runtime gold mappings and per-game generated mappings are not released; anyone who owns a listed game at the listed build can recapture the gold with the released capture scripts and check it against the published hash. The published generated hash is that of the research-configuration map scored in this paper; the released generator, whose spelling table does not use runtime dumps, writes a map that can differ from it in the letter case of a few names, and the released benchmark list gives the SHA-256 of the released generator's map and its score per game, next to the research hash.
For every scored game the generated hash is of the full-pipeline v11 map scored in this paper (re-scoring each written file reproduces its score; the CUE4Parse test of Section~\ref{sec:realuse} used the earlier v10 maps).
Steam serves the current build by default, so the supplementary file also lists each game's depot manifest IDs for retrieving the listed build.
\paragraph{Versions used.}
The released generator is commit \texttt{6fe7f76} of its repository; commit \texttt{7fec886} adds the full-load harness, the load-check patch, the reader fixes as one patch over \texttt{c95dd9b5}, and the scripts that compute the full-load tables; commit \texttt{8b7f990} adds the released generator's map hashes and scores to the benchmark list; commit \texttt{5c37005} replaces the harness's string-based hashing with the streaming version of Section~\ref{sec:readerfix}, which gives the same hashes; commit \texttt{dfa0cb7} adds the scripts that compute the all-121 reload comparison and Table~\ref{tab:fullfix}, and a per-package count of the exports that went through the temporary file.
The full-load evaluation used CUE4Parse at the pinned upstream commit \texttt{7353666}, with a small patch that adds the load checks of Section~\ref{sec:fullload} and changes nothing else; the reader-fix rerun used the same patch on an integration branch based on upstream commit \texttt{c95dd9b5} that merges the fixes of Table~\ref{tab:readerfix}.
\paragraph{Public form of the reference-spelling rule.}
The research rule votes with runtime dumps of the most similar development games, which are not released.
The public tool instead ships a derived spelling table keyed by engine version (major.minor): 16 versions with 140 rows, plus one table per major version pooled over its minor versions (103 rows); about 15\,KB, holding only identifiers and vote counts, with no game names or structures.
A name is respelled when at least two same-version games agree on one other spelling, none kept the generated one, and the spelling occurs in the executable.
The rows are built from the released generator's own output on the development games only, so they describe the generator that ships.
When the executable stores no engine build string, the minor version is estimated from the reflected type names, using a list of type names added or removed in each engine version, built from the development games only. On the 109 of the 121 scored executables that do store the build string, the estimate matches it on 105, and the 4 misses are neighbouring versions. When no estimate is possible, the pooled table for the major version is used.
Anyone can regenerate both files from their own captures with the released build scripts.
All other accuracy numbers in this paper use the research rule.
With the released configuration (development scored with leave-one-out tables and version lists), there are 5,679 wrong fields on development (research rule 5,696), 3,792 on the old hold-out (3,774) and 3,684 on the fresh set (3,707); field accuracy is 99.923\%, 99.885\% and 99.867\%. With the shipped table, which includes each development game's own votes, development has 5,651.
The first release run had found two faults in this step, both fixed before these numbers: games whose executable stores no build string got no spelling votes at all (12 games, 11 of which then lost corrections), and table rows built from an older generator's output respelled names the current generator already gets right (one hold-out game, 102 fields worse).
The per-game totals of that run flagged them; both causes were found and fixed on development data, and no threshold was tuned.
One known residual remains: on one hold-out game the released configuration has 31 more wrong fields than the research rule (413 against 382). There the generator itself emits two names in a capitalised spelling that it gets right on every development game, so the rebuilt table has no row for them, while the research rule's references happened to correct them. We found this only on a hold-out game and deliberately did not chase it, since that would mean tuning on hold-out data; it is recorded as an open generator issue.
Two development games are also worse, by 8 and 10 fields, but only under leave-one-out: three development games share their engine version, so leaving one out leaves two voters. With the shipped table both equal the research rule.
Spellings taken from the public engine source instead were worse than having no rule at all on every split (for example 9,803 against 6,300 wrong development fields).

\paragraph{Cost to reproduce.}
On 22 November 2026 (US Steam store), 127 of the 132 listed games were free (all 24 fresh, all 33 old hold-out and 70 of the 75 development games) and 3 development games were paid, with a total list price of US\$80.97 and the same total at the current price (none was discounted that day); one of them (US\$24.99) is encrypted and not scored, so the scored games cost US\$55.98.
Two development games, app IDs 4553860 and 5090080, were no longer offered on the store, so reproducing them needs an existing copy.

\paragraph{Ethics.}
The tool reads the user's own game files and writes a mapping; it does not execute or modify the game, and its output contains type metadata only, not game content.
Encrypted archives are refused by default; users may supply their own key; the tool does not obtain or break keys. Encrypted benchmark games were excluded, not read.
All benchmark captures ran offline with network access denied, and games with an anti-cheat component were never launched.

\paragraph{Acknowledgements.} The author thanks the maintainers of CUE4Parse, UE4SS, UnrealMappingsDumper, Dumper-7, UEDumper and jmap, whose open-source work made the runtime reference and the real-use evaluation possible. The author used Claude (Anthropic), an AI assistant, to help write code, run experiments and draft text; the author checked all results and takes full responsibility for the content.

% ---------------------------------------------------------------

\appendix
\section{Benchmark list}
\label{app:bench}
All 132 games: 75 development, 33 old hold-out and 24 fresh hold-out.
Build is the Steam build ID of the installed copy that was captured; the executable hash was checked before and after each capture.
Hashes are the first 12 hex digits of the SHA-256; the supplementary file \texttt{benchmark-list.csv} has the full hashes and the depot manifest IDs.
The generated map is the full-pipeline v11 map; encrypted games have no generated map, and for the five UE $<$4.17 games the generator, out of scope there, writes an empty map.
{\footnotesize
\begin{longtable}{lllllll}
\toprule
Split & App ID & Build & UE & Status & Gold SHA-256 & Generated SHA-256\\
\midrule
\endhead
dev & 605610 & 1720434 & 4.13.1 & UE $<$4.17 & \texttt{2da305fe9d56} & \texttt{out of scope}\\
dev & 612590 & 1721433 & 4.13.0 & UE $<$4.17 & \texttt{ea8263f9e532} & \texttt{out of scope}\\
dev & 934860 & 3374034 & 4.19.2 & scored & \texttt{db5b4f2e3c35} & \texttt{9311dff03f11 (v11)}\\
dev & 969220 & 4399109 & 4.16.3 & UE $<$4.17 & \texttt{37bcd3447d1f} & \texttt{out of scope}\\
dev & 1058370 & 4106261 & 4.21.0 & scored & \texttt{8336d3119292} & \texttt{de2cb4bb2a1c (v11)}\\
dev & 1062060 & 4653344 & 4.22.3 & scored & \texttt{8a00f55dabfc} & \texttt{76c0afd0b30a (v11)}\\
dev & 1062070 & 4600507 & 4.22.3 & scored & \texttt{0a9b57ee1c31} & \texttt{825b2c7f4e20 (v11)}\\
dev & 1062080 & 4617280 & 4.22.3 & scored & \texttt{4181d723e7eb} & \texttt{3d3d9788c47b (v11)}\\
dev & 1062100 & 4990762 & 4.23.1 & scored & \texttt{47fd49661d46} & \texttt{7c051a6580d4 (v11)}\\
dev & 1096710 & 3865817 & 4.15.1 & UE $<$4.17 & \texttt{87e8064f1ccb} & \texttt{out of scope}\\
dev & 1175890 & 5239933 & 4.23.1 & scored & \texttt{58f3876d4675} & \texttt{1edc463d108a (v11)}\\
dev & 1214360 & 5297635 & 4.23.0 & scored & \texttt{dc83e46ef86b} & \texttt{fbfd1bc05495 (v11)}\\
dev & 1323410 & 5129218 & 4.22.3 & scored & \texttt{c66aa9c04c00} & \texttt{635a70474970 (v11)}\\
dev & 1324220 & 12570115 & 4.26.2 & encrypted & \texttt{f405ea2abda1} & \texttt{--}\\
dev & 1344810 & 5129297 & 4.22.3 & scored & \texttt{cc4f75a53f8c} & \texttt{1e353fccf06c (v11)}\\
dev & 1363810 & 5308839 & 4.22.3 & scored & \texttt{7a9b7799a765} & \texttt{6694fea3fcac (v11)}\\
dev & 1410830 & 5876727 & 4.25.3 & scored & \texttt{7b094add3d23} & \texttt{2216a11fa2a6 (v11)}\\
dev & 1410840 & 5875807 & 4.25.3 & scored & \texttt{ab1218ca3c1f} & \texttt{8eac7b651e22 (v11)}\\
dev & 1410850 & 6051059 & 4.25.3 & scored & \texttt{41d80fc56f65} & \texttt{35a68c77de11 (v11)}\\
dev & 1410870 & 5876285 & 4.25.3 & scored & \texttt{d77455c297cb} & \texttt{1b0f4c03b8bb (v11)}\\
dev & 1463730 & 7665899 & 4.25.4 & scored & \texttt{4903e06616d4} & \texttt{93e8c7edff33 (v11)}\\
dev & 1490860 & 6951766 & 4.25.4 & scored & \texttt{43e62d7dfb74} & \texttt{2d9c3958269b (v11)}\\
dev & 1491650 & 6942535 & 4.25.4 & scored & \texttt{ae14ff869535} & \texttt{b8160dc9ee8d (v11)}\\
dev & 1497110 & 6998745 & 4.25.4 & scored & \texttt{7e13a0abf327} & \texttt{7956b7ecd886 (v11)}\\
dev & 1545710 & 6658775 & 4.26.2 & encrypted & \texttt{9493d9b6af72} & \texttt{--}\\
dev & 1555710 & 7011170 & 4.25.4 & scored & \texttt{b3f244e47d14} & \texttt{94daeb5b7c5e (v11)}\\
dev & 1702970 & 8393064 & 4.26.2 & encrypted & \texttt{38d85a11d3d3} & \texttt{--}\\
dev & 1702990 & 8311850 & 4.26.2 & scored & \texttt{03ab99b8ac03} & \texttt{a04e37d60274 (v11)}\\
dev & 1761260 & 9094127 & 4.27.0 & scored & \texttt{1f24be6beefb} & \texttt{7ef5954f78ca (v11)}\\
dev & 1768550 & 9004006 & 4.27.0 & scored & \texttt{60e09b185aca} & \texttt{56137fca2645 (v11)}\\
dev & 1903340 & 23765773 & 5.4.4 & scored & \texttt{0149b44cecb2} & \texttt{5f9b5db7287b (v11)}\\
dev & 1903640 & 8685294 & 4.27.2 & scored & \texttt{9d156af0e79b} & \texttt{770c13198251 (v11)}\\
dev & 1992850 & 9103192 & 4.26.2 & scored & \texttt{33d47b19ac5a} & \texttt{4ab57ca135d7 (v11)}\\
dev & 2004680 & 24321921 & 5.7.4 & encrypted & \texttt{1ca604172ca1} & \texttt{--}\\
dev & 2010170 & 8782418 & 4.20.3 & scored & \texttt{b8aeaad88afa} & \texttt{d2ef9c66f398 (v11)}\\
dev & 2142840 & 10353774 & 5.0.3 & scored & \texttt{b09fc8b670b1} & \texttt{cd3b350e3c71 (v11)}\\
dev & 2161880 & 10104934 & 4.27.2 & encrypted & \texttt{c2848504252a} & \texttt{--}\\
dev & 2163100 & 11587626 & 5.1.0 & scored & \texttt{9f8f5592d24c} & \texttt{186e2a752bde (v11)}\\
dev & 2163140 & 11478406 & 5.1.1 & scored & \texttt{d18c36e57636} & \texttt{24dcc555e764 (v11)}\\
dev & 2163210 & 11523098 & 5.1.0 & scored & \texttt{d770ccf0b70e} & \texttt{abad2f3eb158 (v11)}\\
dev & 2322020 & 11190532 & 5.1.1 & scored & \texttt{80bdfb755f0d} & \texttt{8a2fd33846ee (v11)}\\
dev & 2365810 & 13615456 & 5.1.1 & scored & \texttt{bccb02c3866a} & \texttt{37cb13efdb02 (v11)}\\
dev & 2387110 & 10941471 & 4.26.2 & scored & \texttt{65b6f4bc91b4} & \texttt{cff5a782b948 (v11)}\\
dev & 2417620 & 11133592 & 4.26.2 & scored & \texttt{3a50d1d26cd7} & \texttt{d04f1dd8109a (v11)}\\
dev & 2431320 & 11239449 & 5.0.2 & scored & \texttt{f996940d13b2} & \texttt{f0e0a64bb033 (v11)}\\
dev & 2431330 & 11203876 & 5.0.2 & scored & \texttt{ec7d5083196e} & \texttt{82aab818745a (v11)}\\
dev & 2491710 & 13199219 & 5.3.1 & scored & \texttt{97980df45f1a} & \texttt{b79c12b5e48f (v11)}\\
dev & 2491720 & 12934650 & 5.2.1 & scored & \texttt{4483b4363ea8} & \texttt{487cc5607e69 (v11)}\\
dev & 2707340 & 14953859 & 5.4.1 & scored & \texttt{107362fa7708} & \texttt{7d50269c8a1e (v11)}\\
dev & 2707370 & 24958966 & 5.5.4 & scored & \texttt{8d081f253f4b} & \texttt{86686da66fea (v11)}\\
dev & 2707380 & 14925273 & 5.3.2 & scored & \texttt{a00a282cb49f} & \texttt{905984e40e3b (v11)}\\
dev & 2707390 & 14923665 & 5.3.2 & scored & \texttt{710f012d774a} & \texttt{93e5c156df8d (v11)}\\
dev & 2707400 & 14783885 & 5.3.2 & scored & \texttt{540b84f95ec4} & \texttt{a35629e922b5 (v11)}\\
dev & 2707410 & 13093345 & 5.1.1 & scored & \texttt{6a47f9ab3e8a} & \texttt{e11674f1356a (v11)}\\
dev & 2983070 & 14285777 & 5.2.0 & scored & \texttt{52b3743a24dd} & \texttt{2c1bb461306f (v11)}\\
dev & 3042010 & 14630658 & 5.2.1 & scored & \texttt{7d15561d835f} & \texttt{9560053b23bd (v11)}\\
dev & 3126140 & 16476439 & 5.3.2 & scored & \texttt{909c6618c8e5} & \texttt{534bfce5f2cf (v11)}\\
dev & 3200130 & 16762790 & 5.4.4 & scored & \texttt{aae37d727b6f} & \texttt{1b8595c61d84 (v11)}\\
dev & 3200140 & 17294295 & 5.4.4 & scored & \texttt{2ba4f810807f} & \texttt{84548103f37d (v11)}\\
dev & 3224960 & 20984655 & 5.4.4 & scored & \texttt{f638cf9e4a35} & \texttt{f7ce169455bd (v11)}\\
dev & 3365840 & 22218222 & 5.4.1 & scored & \texttt{adbd25a66b77} & \texttt{905651154499 (v11)}\\
dev & 3365850 & 19011847 & 5.4.1 & scored & \texttt{5662ae591314} & \texttt{c1bf10b4745b (v11)}\\
dev & 3365860 & 19008150 & 5.4.1 & scored & \texttt{6de706c90047} & \texttt{41e5476b9a2f (v11)}\\
dev & 3366020 & 16669464 & 5.5.0 & scored & \texttt{ffa4c1b25865} & \texttt{6ae7db2c7bcf (v11)}\\
dev & 3617780 & 18069386 & 5.4.1 & scored & \texttt{d9a58a002037} & \texttt{b3e5f4bb88f2 (v11)}\\
dev & 3640620 & 26097974 & 5.7.4 & scored & \texttt{95856c49cb14} & \texttt{234a59465591 (v11)}\\
dev & 3642710 & 20453304 & 5.5.4 & scored & \texttt{697b3b055460} & \texttt{b2cf23c30ae6 (v11)}\\
dev & 3854930 & 22746466 & 5.6.1 & scored & \texttt{35f888eb4663} & \texttt{a4dcb170a4b0 (v11)}\\
dev & 3854940 & 21195597 & 5.6.1 & scored & \texttt{7fd731884651} & \texttt{ab93f65712dd (v11)}\\
dev & 4319370 & 23184868 & 5.7.4 & scored & \texttt{2fdb26625b48} & \texttt{a588a0fa995a (v11)}\\
dev & 4508350 & 23660981 & 5.6.1 & scored & \texttt{85c9624cd898} & \texttt{4d1b48b62a99 (v11)}\\
dev & 4515400 & 23957349 & 5.6.1 & scored & \texttt{9eedfd451064} & \texttt{cc10b0716a6d (v11)}\\
dev & 4515420 & 23957663 & 5.6.1 & scored & \texttt{1ba512c31169} & \texttt{cb77f0b51443 (v11)}\\
dev & 4553860 & 26050654 & 5.6.1 & scored & \texttt{a96aba865391} & \texttt{a2e426ddc6bd (v11)}\\
dev & 5090080 & 25985740 & 5.6.1 & scored & \texttt{424f71697f3f} & \texttt{040967d959e3 (v11)}\\
hold & 625470 & 2670339 & 4.16.1 & UE $<$4.17 & \texttt{a239ee303ded} & \texttt{out of scope}\\
hold & 934830 & 3362236 & 4.20.2 & scored & \texttt{22607b040399} & \texttt{f73213c19c13 (v11)}\\
hold & 934850 & 3362590 & 4.19.2 & scored & \texttt{e4c0fbc5271b} & \texttt{af3a7d7eca1a (v11)}\\
hold & 1181640 & 5533668 & 4.23.1 & scored & \texttt{7beca6239317} & \texttt{ab4620df2ed9 (v11)}\\
hold & 1346310 & 5137133 & 4.22.3 & scored & \texttt{7d827b408650} & \texttt{2f368432a81e (v11)}\\
hold & 1357710 & 5211081 & 4.20.3 & scored & \texttt{2257399c6ffa} & \texttt{3f6b94e86f33 (v11)}\\
hold & 1410860 & 5895504 & 4.25.3 & scored & \texttt{790d2879a190} & \texttt{f2a9c7099075 (v11)}\\
hold & 1639650 & 6706809 & 4.25.3 & scored & \texttt{03b251f36018} & \texttt{28c6a7d7f24e (v11)}\\
hold & 1702980 & 8254294 & 4.26.2 & scored & \texttt{69789ddfcc7f} & \texttt{181a3cc2a607 (v11)}\\
hold & 1976490 & 8586839 & 4.25.4 & scored & \texttt{efc31467c3b8} & \texttt{360b142fe782 (v11)}\\
hold & 1981300 & 8754399 & 4.26.2 & scored & \texttt{19a16a4e5b53} & \texttt{de5bf4ca683c (v11)}\\
hold & 2098350 & 16836776 & 4.26.2 & scored & \texttt{23ea60aa0b8a} & \texttt{4d1552a10b11 (v11)}\\
hold & 2142850 & 10361641 & 4.27.2 & scored & \texttt{bc6fbbb49749} & \texttt{8d758e39e10a (v11)}\\
hold & 2163220 & 11586934 & 5.1.0 & encrypted & \texttt{88a8da686ac4} & \texttt{--}\\
hold & 2205950 & 9842741 & 4.25.4 & scored & \texttt{49bf87cbf38a} & \texttt{a16937a9ad4e (v11)}\\
hold & 2218950 & 10060270 & 5.0.3 & scored & \texttt{0fca64f4eb39} & \texttt{6f5155d4a2b9 (v11)}\\
hold & 2363730 & 11879371 & 4.26.2 & scored & \texttt{ecc24c2dc20f} & \texttt{4d1552a10b11 (v11)}\\
hold & 2497390 & 11993299 & 5.1.1 & scored & \texttt{7cad8ef9dcc8} & \texttt{cf4c9b8f4ee7 (v11)}\\
hold & 2667430 & 13931099 & 5.1.1 & scored & \texttt{c495aa683a91} & \texttt{513484ae8acf (v11)}\\
hold & 2707360 & 14952479 & 5.3.2 & scored & \texttt{864cf850c5f0} & \texttt{3bf1cfd4a7b7 (v11)}\\
hold & 2830700 & 14522032 & 5.3.2 & scored & \texttt{7b251c528c01} & \texttt{6033032e9853 (v11)}\\
hold & 3365820 & 18966558 & 5.4.4 & scored & \texttt{b0581c4c97c3} & \texttt{9d5185351909 (v11)}\\
hold & 3365900 & 19068607 & 5.4.1 & scored & \texttt{485ae4bef732} & \texttt{bc5de001a9c5 (v11)}\\
hold & 3365910 & 19570747 & 5.5.0 & scored & \texttt{8c081768799d} & \texttt{840746ba70e1 (v11)}\\
hold & 3433290 & 18903480 & 5.0.3 & scored & \texttt{a6a89ac2064d} & \texttt{d703910d7356 (v11)}\\
hold & 3533890 & 19727837 & 5.5.3 & scored & \texttt{7391028546d0} & \texttt{ddd3ff7578ea (v11)}\\
hold & 3573020 & 17584502 & 5.2.1 & scored & \texttt{7d3382da08af} & \texttt{a08e79465a8a (v11)}\\
hold & 3638410 & 19066143 & 5.4.4 & scored & \texttt{57b0ac10660f} & \texttt{2499e6143d08 (v11)}\\
hold & 3642480 & 18807167 & 5.4.1 & scored & \texttt{ea4eb5f07a1d} & \texttt{964ce496746d (v11)}\\
hold & 3689210 & 24174511 & 5.4.4 & scored & \texttt{e4076ee0b789} & \texttt{d99f8e9ec40b (v11)}\\
hold & 3714700 & 18302792 & 5.3.2 & scored & \texttt{c0d8f16e9499} & \texttt{df8ae31d4e98 (v11)}\\
hold & 4515410 & 23624715 & 5.5.4 & scored & \texttt{bc4951705a5d} & \texttt{d853f5143209 (v11)}\\
hold & 4515430 & 23956671 & 5.6.1 & scored & \texttt{dfb47fcdb564} & \texttt{4f042b8c5c74 (v11)}\\
fresh & 1967100 & 8540507 & 4.25.4 & scored & \texttt{8c3a00c75863} & \texttt{79028ec3ab70 (v11)}\\
fresh & 1991570 & 8955870 & 4.26.2 & scored & \texttt{b66121775fd9} & \texttt{ef6e33e27e88 (v11)}\\
fresh & 2073410 & 9024857 & 4.25.4 & scored & \texttt{d0d15b6bd7af} & \texttt{81ae204a55b2 (v11)}\\
fresh & 2431310 & 11687124 & 5.0.2 & scored & \texttt{23c7d53056cc} & \texttt{a7929bed0bb5 (v11)}\\
fresh & 2439100 & 12123518 & 4.26.2 & scored & \texttt{967582335063} & \texttt{2f0206ea96df (v11)}\\
fresh & 2456320 & 11813758 & 4.27.2 & scored & \texttt{6cb4ad1255db} & \texttt{e221c974dd48 (v11)}\\
fresh & 2686740 & 12664634 & 4.26.2 & scored & \texttt{adf3b559688c} & \texttt{285b98789175 (v11)}\\
fresh & 2730030 & 12894328 & 4.26.2 & scored & \texttt{067adc322a6b} & \texttt{dcff8b19e76f (v11)}\\
fresh & 2858380 & 14992387 & 5.2.1 & scored & \texttt{b3f62ee75ab9} & \texttt{43dcb22f290c (v11)}\\
fresh & 2878180 & 14164102 & 5.2.1 & scored & \texttt{9be1e1ac3929} & \texttt{f9d7e1c6c55e (v11)}\\
fresh & 2916580 & 16108614 & 5.2.1 & scored & \texttt{3b1d581ed2bd} & \texttt{2f6b1a3c4acf (v11)}\\
fresh & 2942810 & 15057512 & 5.0.3 & scored & \texttt{ca85404d2108} & \texttt{bbed813bcdf0 (v11)}\\
fresh & 2983040 & 14285534 & 5.2.0 & scored & \texttt{7b08040dd63f} & \texttt{531f5ff21653 (v11)}\\
fresh & 3041920 & 14831386 & 5.2.0 & scored & \texttt{a4b8aa31c3a9} & \texttt{0567577fc148 (v11)}\\
fresh & 3189180 & 17615141 & 5.3.2 & scored & \texttt{32b271f53c4c} & \texttt{7be93f853757 (v11)}\\
fresh & 3290070 & 16597632 & 5.4.4 & scored & \texttt{9debf3fa128b} & \texttt{9d9758f64794 (v11)}\\
fresh & 3560620 & 17626938 & 5.2.1 & scored & \texttt{84ffb60290e0} & \texttt{2084607a4d84 (v11)}\\
fresh & 3814650 & 19292118 & 5.4.4 & scored & \texttt{c4f137ee3c6b} & \texttt{32ca8cf7cc19 (v11)}\\
fresh & 3941140 & 20550908 & 5.4.4 & scored & \texttt{bfb4d6047eca} & \texttt{2c0da5ded933 (v11)}\\
fresh & 4011870 & 20039713 & 5.2.1 & scored & \texttt{5f6d4c733f92} & \texttt{f6afeb6299a6 (v11)}\\
fresh & 4395790 & 23733740 & 5.4.4 & scored & \texttt{9406748494a1} & \texttt{faafa8ee4906 (v11)}\\
fresh & 4440710 & 25244433 & 5.7.4 & scored & \texttt{820d253b4edd} & \texttt{2db49ccccf09 (v11)}\\
fresh & 4598370 & 23088450 & 5.6.1 & scored & \texttt{6a40cefe5cce} & \texttt{044d25e2858b (v11)}\\
fresh & 4897740 & 25382383 & 5.4.4 & scored & \texttt{6ea55e5354c4} & \texttt{4cad26f6d584 (v11)}\\
\bottomrule
\end{longtable}}

\end{document}